# Parallel Algorithm for Longest Common Subsequence in a String

TIRTHARAJ DASH

Department of Information Technology
National Institute of Science and Technology
Berhampur-761008, India
Email: tirtharajnist446@gmail.com

TANISTHA NAYAK

Department of Information Technology
National Institute of Science and Technology
Berhampur-761008, India
Email: tanisthanist213@gmail.com

*Abstract*— **In the area of Pattern Recognition and Matching, finding a Longest Common Subsequence plays an important role. In this paper, we have proposed one algorithm based on parallel computation. We have used OpenMP API package as middleware to send the data to different processors. We have tested our algorithm in a system having four processors and 2 GB physical memory. The best result showed that the parallel algorithm increases the performance (speed of computation) by 3.22.**

*Keywords-Pattern Recognition; Matching; Longest Common Subswquence; OpenMP; Speed of Computation;*

## I. INTRODUCTION

The longest common subsequence (LCS) of two strings is one of the interesting areas in combinational pattern matching. The LCS problem is related DNA or protein alignment, file compression, speech recognition, FPGA circuit minimization and bioinformatics. In biological applications, we want to compare the DNA of two (or more) different organisms. A strand of DNA consists of a string of molecules called **bases**, where the possible bases are adenine, guanine, cytosine and thymine. Representing each of these bases by their initial letters, a strand of DNA can b expressed as a string over the finite set {A, C, G, T}.
For an example the DNA of one organism may be
$$S_1 = \{ACCGGT……GCGAA\},$$
While DNA of another string may be
$$S_2 = \{GTCGTTCG…..TGGTACAA\}.$$
Our goal of comparing two strands of DNA is to determine how "similar" the two strands are, as some measure of how closely related the two organisms are. Similarly can be and is defined in many different ways [1]. However we can say the two strings are similar if one is substring of other. Alternatively we can say that two strands are similar if the number of changes needed to turn one into another is small. The Longest Common Problem seeks the longest subsequence of every number of a given set of strings. For more than two input strings, and the existing exact solutions are impractical for large input sizes which give rise to NP completeness problem [1,2]. The Shortest Common Subsequence (SCS) and the Longest Common Subsequence (LCS) are classical problems in computer science. Longest Common Subsequence (LCS) comes under Dynamic Programming which is similar to Divide and Conquer method [3,4]. It solves the problem by combining the solutions to sub problems. In case of Dynamic problem, the sub problems are not independent i.e. sub-problem share sub-problem. When the sub-problems are not independent, in that case we can use divide and conquer method. The Dynamic programming algorithms solve every sub-problem only once and then save the answer in a table, thereby the work of re-computing the answer every time the sub-problem is encountered [5,6]. The Dynamic Programming technique is used for optimization Problems. In such problems there can be many possible solutions, where we wish to find the "best" way of doing something. One way to solve optimization problem is to try all the possible solutions and then pick out the best. But here the time and space requirements are exponential in nature. So, by Dynamic technique we can drastically reduces the amount of enumeration by avoiding the enumeration of some decision sequences that cannot be possibly be optimal. In the Longest Common Subsequence Problem, we are given two sequences $X = <x_1, x_2 … x_n>$ and $Y= <y_1, y_2 … y_n>$ and wish to find a maximum length common subsequence of X and Y. One way to solve the longest common subsequence problem is to enumerate all subsequence of x and take the largest one that is also a subsequence of Y. Since each character of X is either in or not in a subsequence there are potentially $2^m$ two different subsequence of X, each of which requires O (n) time to determine whether it is a subsequence of Y. Thus the brute force approach yields exponential algorithm that runs in O $(2^m.n)$ time, which is very inefficient [2,7].

Here in this section, we have discuss how to use dynamic programming to solve the longest common subsequence problem which is faster than brute force method.

### A. Characterizing a longer common subsequence

A brute force approach to solving the LCS problem is to enumerate all sequences of X and check each subsequence to see if it is also subsequences to see if it is also subsequences of Y, keeping track of the longest subsequences found. The LCS problem has an optimal substructure property in the following theorem. The optimal substructure can be described as follows [2].

Let X = <$x_1$, $x_2$ . . . $x_n$> and Y= <$y_1$, $y_2$ . . . $y_n$> be the sequences, and let Z =($Z_1$,$Z_2$,. . . . . . . . . ., $Z_k$ ) be any LCS of X and Y.

1. If $x_m$ = $y_n$, then $Z_k$ = $X_m$ =$Y_n$ and $Z_{k-1}$ is LCS of $X_{m-1}$ and $Y_{n-1}$.

2. If $x_m \neq y_n$, then $Z_k \neq X_m$ implies that $Z_{k-1}$ is an LCS of $X_{m-1}$ and Y.

3. If $x_m \neq y_n$, then $Z_k \neq y_n$ implies that Z is an LCS of X and $Y_{n-1}$.

### B. A recursive solution

There are either one or two sub-problems to study when finding an LCS of X = <$x_1$, $x_2$ . . . $x_n$> and Y= <$y_1$, $y_2$ . . . $y_n$> . This means that to find the LCS of X AND Y.
If $x_m$ = $y_n$, then find LCS of $Xm_{-1}$ and $Y_{n-1}$ and append $x_m$ = $y_n$ to this get the LCS of X and Y [2].
If $x_m \neq y_n$, (a) Find the LCS of $X_{m-1}$ and Y. (b) Find the LCS X and $Y_{n-1}$. And take the larger of (a) and (b), and that is solution of the LCS problem where as in the case $x_m \neq y_n$, we have to solve two sub problems, i.e. to solve $X_{m-1}$ and Y, and X and $Y_{n-1}$. Thus we start with small problem, find LCS and grow our solution. Let us define c[i, j] to be length of and grow our solution [2].

### C. Computing the length of an LCS

In this step, we can write an exponential time recursive algorithm to compute the length of an LCS of two sequences based on the recursive formula which is described in equation (1). Science there are only Θ (m n) distinct sub-problems, however we can use dynamic programming to compute the solution bottom up. The procedure LCS –length takes two sequences = <$x_1$, $x_2$ . . . $x_n$> and Y= <$y_1$, $y_2$ . . . $y_n$> as inputs. I t stores the c[i, j] values in a table c[0…m, 0…n] whose entries are compared in row major order. It also maintains the table b[1…m, 1…n] to simplify construction of optimal solution. The b[i, j] points to the table entry corresponding the optimal solution chosen when computing c[i,j]. The procedure returns the b and c tables; c[m, n] contains the length of an LCS of X and Y. LCS works on following algorithms. The algorithms have been given below [2].

```
LCS-length (x, y)
    1.    Calculate the length of x.
    2.    Calculate the length of Y.
    3.    For i = 1 to length of m
    4.        set C[i, 0] =0;
    5.    For J = 1 to length of n
    6.        Set C[0,j] =0;
    7.    If (xi = yj)
    8.        then c[I, j] = c[i-1,j-1] +1 and b[i, j]
            = "◆".
    9. Else if (c[i -1, j]) ≥c[i, j-1],
    10.       then c[i,j] =c[i-1,j] and b[i, j] = "↑"
    11. Else c[i, j] = c[I, j-1], then b[i, j] = "←"
    12. Return value of c and b.
```

### D. Constructing an LCS

The table b returned by LCS length can be used to quickly to construct an LCS of X = <$x_1$, $x_2$ . . . $x_n$> and Y= <$y_1$, $y_2$ . . . $y_n$>. Let us begin at b[m, n] and trace through the table following the arrows. Whenever we encounter a "◆" in entry b[I, j], it implies that xi =yi is an element of the LCS. The elements of the LCS are encountered in reversed order by this method. Then the following algorithms print LCS in forward order [2].

```
Print-LCS (b, X, i, j)
{
   if (I =0 or j =0)
    Then return;

   if(b[I ,j]= "◆")
    Then Print-LCS (b, X, i-1, j-1);
       Print Xi ;

   Else if (b[I ,j]) = " ↑ "
    then Print-LCS (b, X, i-1, j-1);
    then Print-LCS (b, X, i, j-1);
}
.
```

## II. PARALLEL ALGORITHM

We have designed a parallel algorithm to find the longest common subsequence (LCS) in a string. The algorithm is based on dynamic programming concept. The process can be applied to taking decision of genetics related problems in minimal time of computation. In the whole course of designing process we considered that we are considering DNA strings which contains A-T-G-C links [4,7,8]. There are two DNAs to be considered; (i) parent and (ii) child. The goal is to know whether the child is having matching characteristics like the parent or not; and if yes how much percentage. We can also consider two patterns also for pattern matching problems. [9]

The designed algorithm is a block-stripped based algorithm, where the speed up increases with length of the string.

### A. Proposed Algorithm

*Inputs:*
Let's consider two strings Parent (P) and Chile (C).
P = < ATGCCCCAA …….. up to length M >
C = < ATGGGGGCA ……. up to length N >

SYM = Symbol matrix of size M×N
WT = Weight matrix of size M×N

Np = Number of Processors of the system; this information can be obtained by using the OpenMP function *omp_get_num_threads* ().

EXEC_TIME = Time of Execution of the algorithm; to know the time *omp_get_wtime* () is used. The statement where it is first called is the start of timer.

*Output:* Time of Computation and Length of LCS

1. Send first request to root processor to identify itself [using *omp_get_thread_num( )*]
2. *if* Root equals to 0 *then*
3. M = Get the length of P;
4. N = Get the length of C;
5. Length = 0; /* length of LCS */
6. EXEC_TIME = *omp_get_wtime ();*
7. *Distribute* the work to NP number of processors for assigning values to SYM and WT
   /* parallel section */
8. Root sends the block section (i) of the matrix to Np number of processors.
9. *endif*
10. *If* Processor Number $\neq$ 0 *then*
11. *Receive* the data from Root processor
12. each processor does the following operations 13-16
13. *if* P[i] equals C[i] *then*
14. *Update* SYM and WT matrix
15. Length = length+1;
16. *Send* SYM, WT and length to Root
17. *endif*
18. *if* Processor Number equals 0 *then*
19. *Receive* and Gather data from all the processors
20. *endif*
21. EXEC_TIME = *omp_get_wtime( )*-EXEC_TIME;
22. *Return* length, EXEC_TIME

### B. Parallel Implentation of the algorithm

The proposed parallel algorithm has been implemented in 'C' language. However, for the parallel implementation we have used an API package called as OpenMP (*www.opemmp.org*). This API package is a middleware which distributes the work to number of processors. But, for the distribution of work, this middleware program has to be instructed by the developer of the programmer. Our contribution in this paper is to compute in parallel the longest common subsequence in a mother string of length N.

The program uses different OpenMP directives to do the work of distribution. One example is given below.

```
#pragma omp parallel
{
      for ( … )
      {
              //instructions to be parallelizable
      }
}
```

### C. Analysis of the developed algorithm

Analysis of the algorithm refers to two things,
1. Space complexity
2. Time complexity

1. *Space complexity:*

As we have considered the lengths of the two strings as M and N, the SYM and WT matrix can be stored in M×N dimensional size. Basically, WT is a matrix of integers and SYM is a matrix of character. So the space complexity of this algorithm is $\Theta(MN)$. However, the parallel execution does not affect the space complexity.

2. *Time Complexity:*

In serial execution, the algorithm takes $\Theta(MN)$ time to compute the result. But, the parallel algorithm distributes the algorithm to Np number of processors. So, the algorithmic time complexity becomes $\Theta\left(\dfrac{MN}{Np}\right)$.

After, the implementation is carried out, two important parameters decide performance.
1. Speed up
2. Scalability

### D. Specification of the system

The system in which the parallel algorithm was implemented and executed was having following specification.

*OS* : **Linux (Ubuntu 11.10)**
*Compiler* : **gcc**
*Memory Size (RAM)* : **2 GB**
*Number of Processors* : **4**

The result section describes the implemented result and the above two parameters.

### III. RESULT AND DISCUSSION

The Table-1 gives the detailed result of the implementation.

**Table-1:** Detailed result of Parallel LCS algorithm

| Length of Parent String (M) | Length of Child String (N) | Execution Time (seconds) | Speed up |
|---|---|---|---|
| 10 | 5 | 0.005 | 0.0 |
| 100 | 10 | 0.47 | 0.78 |
| 200 | 30 | 2.03 | 1.32 |
| 500 | 80 | 4.12 | 2.09 |
| 800 | 100 | 5.34 | 2.34 |
| 1000 | 150 | 7.09 | 3.01 |
| 2000 | 200 | 10.03 | 3.13 |
| *5000* | *200* | *18.90* | *3.22* |
| 5000 | 400 | 24.08 | 3.17 |
| 5000 | 800 | 31.67 | 3.23 |
| 5000 | 1000 | 33.37 | 3.02 |
| 10000 | 1000 | 51.11 | 2.97 |
| 10000 | 1500 | 57.99 | 2.89 |

From the practical implementation, from Table-1, it can be seen that highest speed up could be achieved with M=5000 and N=200. A plot has been given in Figure-1 below to show the timing variation between serial and parallel implementation of the LCS algorithm (speed up curve).

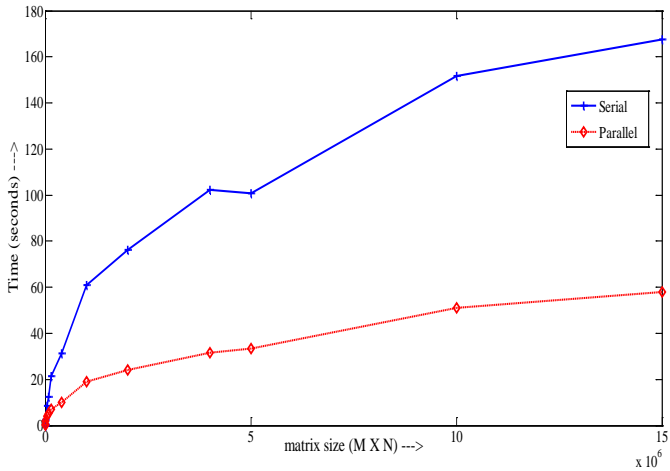

**Figure-1:** plot showing Execution time difference of serial and parallel implementation (speed up curve)

In the plot X-axis is the matrix size (M×N). Figure-2 shows the scalability curve for the proposed parallel method.

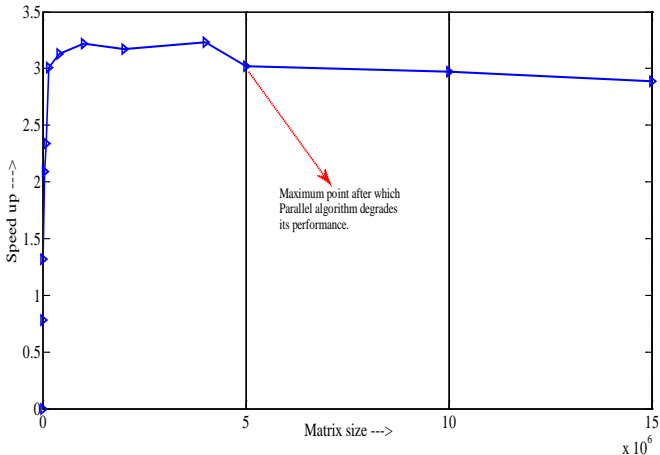

**Figure-2:** Scalability curve for the proposed parallel algorithm showing a maximum input point after which the parallel algorithm degrades its performance.

## IV. CONCLUSION

In this work, we designed a parallel algorithm to find out the Longest Common Subsequence (LCS) in a parent string of length of M. After implementation of the designed parallel algorithm in 'C', the result showed that the developed algorithm is speeding up the computation by Np times; where Np is the number of the processor of the system. In real world applications, this algorithm can be applied to DNA matching and pattern matching techniques.